# JCTC: A Large Job posting Corpus for Text Classification

**Haoyu Xu**

Shanghai Advanced Research Institute, Chinese Academy of Sciences, China
University of Chinese Academy of Sciences, China

**Chongyang Gu**

Shanghai Advanced Research Institute, Chinese Academy of Sciences, China
Department of Communication & Information Engineering, Shanghai University, China

**Han Zhou**

Shanghai Advanced Research Institute, Chinese Academy of Sciences, China

**Sengpan Kou**

Department of Mathematics, The Chinese University of Hong Kong, Hong Kong SAR

**Junjie Zhang**

Department of Communication & Information Engineering, Shanghai University, China

*Abstract*
*The absence of an appropriate text classification corpus makes the massive amount of online job information unusable for labor market analysis. This paper presents JCTC, a large job posting corpus for text classification. In JCTC construction framework, a formal specification issued by the Chinese central government is chosen as the classification standard. The unsupervised learning (WE-cos), supervised learning algorithm (SVM) and human judgements are all used in the construction process. JCTC has 102581 online job postings distributed in 465 categories. The method proposed here can not only ameliorate the high demands on people's skill and knowledge, but reduce the subjective influences as well. Besides, the method is not limited in Chinese. We benchmark five state-of-the-art deep learning approaches on JCTC providing baseline results for future studies. JCTC might be the first job posting corpus for text classification and the largest one in Chinese. With the help of JCTC, related organizations are able to monitor, analyze and predict the labor market in a comprehensive, accurate and timely manner.*

## 1. Introduction

Recruitment websites are the major sources of labor market information. According to a survey conducted by Tencent in 2015[1], 53.71% of job seekers found jobs on the Internet, and 57.61% of job seekers released job-seeking information on recruitment websites. Therefore, a massive amount of recruitment data has been accumulated.

It seems feasible that analysing these online recruitment data could help investigate the situation of labor market as well as the healthiness of national economy. However, the absence of a well-established job title classification framework may induce inaccuracies in collecting statistics. Each individual company chooses the job titles based on their own understandings. Consequently, jobs of similar nature could be in various titles. Moreover, job descriptions sharing the same title may represent different nature of jobs.

Table 1 shows three job postings released by different companies. Each job posting is composed of a job title and description. Posting No.1 and No.2 have the same job title but dissimilar job descriptions. Posting No.2 and No.3 shows resemblance in job descriptions but the two jobs are in different titles. This phenomenon causes job statistics online to be inaccurate, from which it seems impossible to acquire profound insights for labor market analysis

There are some job databases in English, yet these databases also do not have a uniform job title definition. For example, HiQ Labs specializes in HR analytics and has compiled a large dataset of professional profiles from various public online sources. This dataset contains summaries of approximately 25k people as well as the job titles assigned by its employees.

*Corresponding author:*
*Haoyu Xu, Chinese Academy of Sciences, Shanghai Advanced Research Institute, China*
*Email: xuhy@sari.ac.cn*

Job titles are still chosen manually rather than adopting a systematic scheme [2]. Paper [3] is just a simple collection of Internet recruitment data without any classifications by job nature.

*Table 1. Samples of online job postings.*

| No. | Job Title | Job Descriptions (English) | Job Descriptions (Chinese) |
| --- | --- | --- | --- |
| 1 | Data Analyst | Responsibilities: 1. Implement and follow-up promotional events; 2. Take photographs for promotional purposes; 3. Propose recommendations for improving the execution of activities. | 岗位职责:1.落实,跟进促销活动的实施;2.宣传方面的拍照;3.提出改善活动执行的建议。 |
| 2 | Data Analyst | Job description: 1. Possess thorough understanding of mainstream big data products, data exchange, current situation and development trends of processing products; 2. Understand data mining technology, need to have in-depth experience in business intelligence or data mining; 3. Understand distributed data technologies such as Hadoop, HDFS, Hive, etc. | 职位描述:1.对主流大数据产品,数据交换,处理产品的现状和发展趋势有深入了解 2.了解数据挖掘技术,需要有深入的商业智能或者 数据挖掘工作经验 3.了解 Hadoop, HDFS, Hive 等分布式大数据技术. |
| 3 | Big Data Development Engineer | Responsibility: 1. Develop an R & D big data processing framework; 2. Analyse mined data and make prediction for large data platform; 3. Design, research, develop, maintain and continuously optimize the distributed system. | 职责描述:1.负责大数据处理框架的研发设计工作;2.负责大数据平台的海量数据分析挖掘,预测;3.负责分布式系统的设计,研发,维护全过程,系统的不断优化. |

This paper introduces JCTC, a large Job posting Corpus for Text Classification. Firstly, a formal job category system with detailed job definitions is chosen for organizing JCTC, as in the case of Imagenet which uses WorldNet to classify its data [4]. The paper chooses *People's Republic of China Grand Classification of Occupations* (CGCO) [5] as the job classification standard. The CGCO contains all occupations in China and is classified hierarchically according to the job natures. The CGCO is believed to be authoritative because it is issued by Chinese Central Government. Secondly, JCTC is constructed by mapping online job postings to a certain category of the CGCO. For instance, posting No. 2 and No.3 in Table 1 are mapped to "Data analysis and processing engineering technicians" in the CGCO, while posting No.1 is mapped to "Marketing professionals". The construction process is completed by joint machine learning algorithms and human judgements, iteratively. Lastly, several state-of-the-art deep learning classification approaches are implemented on JCTC serving as the baseline performance.

In summary, this paper makes the following two contributions. 1) It proposes a large corpus called JCTC. As far as we know, it is the first job corpus which maps the online job data to a well-established classification system, namely, the CGCO. The choice of the CGCO as the job classification standard facilitates the analysis for labour market, especially for some government authorities. Moreover, to the best of authors' knowledge, JCTC is the largest Chinese corpus for text classification. 2) A novel method to build the corpus is presented. Unsupervised learning algorithm was firstly applied to classify job postings into the CGCO's categories. Subsequently, human judgements were used to determine whether the classification label was correct or not. The human judgement process is superior to other existing methods. Those methods directly classify the data to a specific category, which requires considerable expertise. By contrast, the proposed method reduces the influence of human subjectivity and improves the accuracy rate. The method to construct JCTC is not Chinese specific and most countries also have documents similar to the CGCO. Therefore, the idea proposed in this paper can also be applied to other countries and languages.

The remaining sections of this paper are organized as follows. Section 2 reviews related work on both corpus and construction method. Section 3 presents the framework of constructing JCTC. Section 4 describes the experimental results. JCTC is then discussed in Section 5, followed by the conclusion in Section 6.

## 2. Related work

This section discusses related works from two perspectives, the corpus itself and its construction method. As far as we know, there is not any text classification corpus for job market. Job databases in English do not have a unified job title definition. Rather, job titles were assigned by users empirically[2] [3]. These databases are more like a collection of job postings rather than a corpus for further research and development.

There are several corpuses for text classification in Chinese. However, their sizes are relatively limited. Chinese news corpus released by Institute of automation of Chinese Academy of Science has only 16280 samples [6]. The Fudan news corpus contains 9833 samples for testing and 9804 samples for training [7]. TanCorpV1.0. [8] collected by Tan Songbo et al. manually, which contains 14150 texts.

The absence of the text classification corpus with a considerable size is perhaps due to the lack of reasonable construction schemes. The most popular method to construct a corpus is to crawl the data with tagged correct labels from the Internet. 20 Newsgroups were built from Usenet [9] with each news tagged as sports, politics and economics etc. Most of the sentiment classification corpuses collected user reviews with "like" or "unlike" labels from large web sites, such as [10] [11] [12] [13]. For the case in this paper, the online job postings do have job titles. But the job titles cannot be used as labels because of the arbitrary jot title choice.

While for data without correct labels, it usually involves manual annotation method. The method is well known for its labor intensity and high demands on people's knowledge. Sriram et al. provided a corpus composed of 5407 tweets manually labelled with 5 categories [14]. Dumais et al. provided a collection consisting of 370597 unique pages that were manually classified by trained professional web editors. It had been manually classified into a hierarchy of categories by trained professional web editors. [15]. Although Reuters Corpus Volume I (RCV1) was first classified by auto coding, it also required manual editing and manual correction. A total of 312,140 stories had been reviewed by a second editor in the holding queue, and 13.4% had codes changed by that second editor [16]. Therefore, it seems arduous to try to fulfil the goals of job posting corpus construction by only text clustering or manual processing.

## 3. Corpus construction

Two sets of data are required in constructing JCTC, namely, online job postings and the government-issued CGCO. Online job postings are composed of job IDs, job titles and job descriptions as shown in Table 2. The CGCO is a job classification standard containing job category codes, job category labels and job category descriptions. The task is to assign a job category label to each job posting based on the similarity between the online job description and the CGCO's job category description.

This section describes online job postings, the CGCO, the process of corpus construction and benchmark classification approaches.

### 3.1. Online job postings

The job postings used in this paper are collected from ChinaHR[17], Zhaopin[18] and 51job[19], three of the largest online recruitment websites in China. The original data contains 6447752 job postings from 2536 China's listed companies released from August 2014 to August 2015. Removing the garbled data and those with extremely similar descriptions, 107,328 samples were retained.

### 3.2. CGCO

This paper chooses the CGCO as the job classification standard because of its authority and hierarchy. CGCO was issued jointly by the People's Republic of China Ministry of Labor and Social Security, the State Administration of Quality Supervision, and the National Bureau of Statistics in 2015. Many consultant firms and government authorities use the CGCO as the baseline for labor market analysis.

The CGCO contains all occupations in Chinese, which are organized in a hierarchical way according to the occupations' nature. It has a total of 1481 categories and is organized into a 4-level hierarchy. For example, the professionals & technical taxonomy is structured as follows:

2:Professionals & Technical.
2-02:Engineering technical personnel.
2-02-10:Information and communications engineering and technical personnel.
2-02-10-03:Software engineer.

Please be note that category in this paper means the forth-level category in the CGCO by default.

As shown in Table 2, there are detailed descriptions of each category in the fourth level of the CGCO. Although the description seems to be enough for being differentiated, the CGCO still has some confusing categories, such as category 2-02-10-03 and category 4-04-05-01. Therefore, even experts cannot accurately classify all job postings directly to certain categories. To reduce the influence of human subjectivity in this paper, people are only required to judge whether the job labels assigned by machine learning algorithms are correct or not.

*Table 2. Samples of categories in the CGCO.*

| Category Coding | Category Label | Category Descriptions (English) | Category Descriptions (Chinese) |
|---|---|---|---|
| 2-02-10-03 | Software Engineer | Engineering and technical personnel for computer software research, requirements analysis, design, testing, maintenance and management. Main tasks: 1.Research and apply computer software development technologies and methods; 2.Analyse project or product requirements, write specifications and software design documentation; 3.Design, develop and test computer software; 4.Deployment and integration of computer software; 5.Write and manage software development documents; | 从事计算机软件研究、需求分析、设计、测试、维护和管理的工程技术人员。主要工作任务：1.研究、应用计算机软件开发技术和方法；2.分析项目或产品需求，编写需求说明书及软件设计文档；3.设计、编码和测试计算机软件；4.部署和集成计算机软件；5.编写和管理软件开发文档 |
| 4-04-05-01 | Computer programmer | Persons engaged in computer and mobile programming. Main tasks: 1.Analyse development requirements; 2.Write and submit module design details; 3.Write and modify codes; 4.Verify the correctness of codes and realize module functions. | 从事计算机和移动终端应用程序设计、编制工作的人员．主要工作任务：1．分析开发需求的概要和细节；2．编写、提交模块设计详细文档；3．编写、修改程序代码；4．验证程序代码的正确性和模块功能的实现程度 |

## 3.3. The process of corpus construction

There are two modules in constructing a corpus. In module one, the preliminary corpus was built by the coordination between an unsupervised learning approach and human judgements. In module two, the final corpus was constructed by the teamwork of a supervised learning method and human judgements. The detailed process is described as follows.

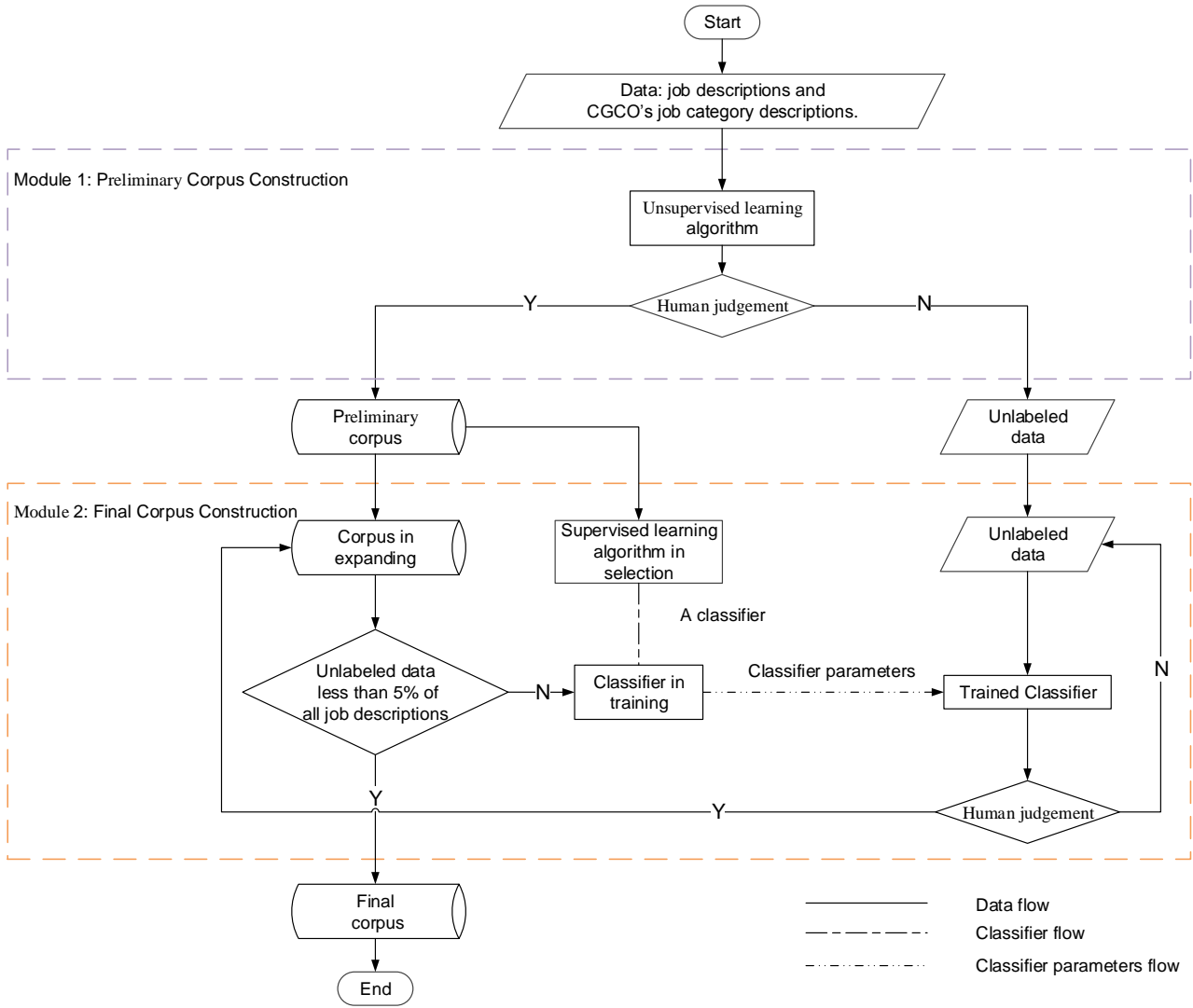

Fig 1. The process of corpus construction

The responsibility of module 1 is to construct the preliminary corpus. The inputs to the module 1 were job descriptions from the recruitment websites and job category labels/descriptions from the CGCO. An unsupervised learning algorithm was used to compute the similarities between each job description and 1481 CGCO's job category descriptions. The category label with the largest similarity value was regarded as the label of the job description. Then, five college students would judge whether the label was correct. If the majorities believed the label given by the computer was correct, the job description would be assigned by the label and added into the preliminary corpus. Otherwise, the job description would be allocated to the set named as unlabelled data.

The responsibility of module 2 is to construct the final corpus, JCTC. The process here was an iterative loop with the preliminary corpus and the unlabelled dataset as inputs. A supervised learning algorithm for classification was firstly determined using the preliminary corpus as the training data set. Then, the category labels were assigned to job descriptions from the unlabelled dataset by the classifier. Those labels unanimously agreed by the learning algorithm and human judgement continued to be added into the corpus. The loop will stop when the size of the unlabelled data was less than 5% of job description inputs. The final corpus is JCTC. .

In the above two modules, the selection of machine learning algorithm directly affects the quality of corpus. The unsupervised learning algorithm and supervised learning algorithm used in this paper would be introduced in section 3.3.1 and 3.3.2, respectively.

*3.3.1. Unsupervised learning algorithm*

The unsupervised learning algorithm used in this paper is <u>W</u>ord <u>E</u>mbedding <u>Cos</u>ine, or WE-cos for short. It is a new method for text similarity calculation proposed by the authors [20].

The novelty of WE-cos is that it considers both the semantic relationship between the vocabularies and the distribution of vocabularies in the corpus. These correlations capture the semantic similarity of words, which overcome the independent assumption of elements in the traditional vector space model. Word embedding is a distributed representation of a word that captures a large number of precise syntactic and semantic word relationships [21] [22][23]. In this work, the Skip-gram model was used to pre-train the word embedding. This model is state-of-the-art in many NLP tasks [24] [25] [26]. The distribution property of the corpus is computed by TF-IDF [27].

The formula for similarity between input texts $D_i$ and $D_j$ is shown in equation (1), where $H_{j,k}$ is the semantic similarity between two word embeddings as calculated by the cosine function. $T(w_g)$ is the TF-IDF value of the word $w_g$ to compute its distribution in the corpus. The parameter p, a real number from 0 to 1, is used to mitigate the influence of text length. The pseudo code of the algorithm is shown in the Figure 2:

$$\text{Sim}(D_i, D_j) = \frac{\sum_{g \in D_i} \sum_{k \in D_j} H_{g,k} \cdot T(w_g) \cdot T(w_k)}{\sqrt{\sum_{g \in D_i} \sum_{l \in D_i} H_{g,l} \cdot T(w_g) \cdot T(w_l)} \cdot \sqrt{\sum_{k \in D_j} \sum_{m \in D_j} H_{k,m} \cdot T(w_k) \cdot T(w_m)}} \quad H > p \quad (1)$$

```
Algorithm Sim(D_i, D_j)
Input     D_i = D_i(T(w_i1), T(w_i2), ···T(w_ig))
          D_j = D_j(T(w_j1), T(w_j2), ···T(w_jk))
          Words: word vector sets
Output    The similarity between the input texts D_i and D_j
Parameter H_{g,k} The semantic similarity between two word embeddings
          p    A real number from 0 to 1
for D_i do
   for D_j do
      H_{g,k} = Cosine(Words[w_ig], Words[w_jk])
      if H_{g,k} > p then
          mul_ij += H_{n,m} * T(w_ig) * T(w_jk);
      end if
   end for
end for
Calculation of mul_i and mul_j are same as mul_ij
return Sim(D_i, D_j) = mul_ij/( Math.sqrt(mul_i) * Math.sqrt(mul_j))
```

Fig 2. The pseudo code of WE-cos

According to [20], the WE-cos outperformed the other three popular methods in F1 value and accuracy [28] evaluation criteria. The F1 value of WE-cos was 9.23% larger than the value for the cosine similarity algorithm [29].

*3.3.2. Supervised learning algorithm*

Support vector machine (SVM) [27] [30] and random forest (RF) [31] are used for classification method selection.

The basic idea of SVM is to solve for a hyper-plane that can correctly partition the training set and maximize the geometric interval. One advantage of the SVM classifier is that linearly inseparable data can be transferred into a higher dimensional space which is linearly separable with an appropriate kernel.

This paper uses the Gauss kernel function and soft interval maximization to train the classification decision function as follows:

$$f(x) = \text{sign}(\sum_{i=1}^{N} \alpha_i^* y_i K(x, x_i) + b^*) \quad (4)$$

where N is the number of samples, y = {−1, +1} is the category of samples. $a^*$ is the weight vector, and $b^*$ means bias vector, sign(x) is a sign function as in equation (5) and k(x, z) is a Gauss kernel function as in equation (6):

$$\text{sign}(x) = \begin{cases} +1, & x \geq 0 \\ -1, & x < 0 \end{cases} \quad (5)$$

$$K(x, z) = \exp\left(-\frac{\|x-z\|^2}{2\sigma^2}\right) \quad (6)$$

, where σ is the width parameter.

RF is an ensemble learning method based on decision trees. First, RF uses random sampling with replacement to get a subspace of features and to construct branches of decision trees. Then, the training data are randomly sampled to generate each tree. Finally, an RF classification model is created by combining all individual trees. For a new sample, each tree can generate a classification label, and the label with the most votes is selected as the final outcome [31].

We compared SVM with RF on the preliminary corpus to determine which classification method is more appropriate. The detailed realization of these steps is illustrated in Section 4.

### 3.4. Benchmark Approaches

An important value of a corpus is its availability of published benchmark results [16]. Good benchmark results serve to ensure that superior new methods are not being compared to artificially low baselines. Therefore, we benchmark several methods that were widely studied in previous text classification experiments.

(1) Three convolutional neural network (CNN) models with different word vectors were selected for the baseline [32]. These CNN architectures may have only one convolutional layer. Two of these models used pre-trained word vectors as inputs. CNN-static used the fixed word vectors while word vectors of CNN-non-static were 'tuned' for a specific task. The other model was called as CNN-rand, which used randomly initialized word vectors as input.

(2) A long-short term memory (LSTM) method was also implemented for the baseline [33]. LSTM refined the recurrent neural network (RNN) architecture around special 'memory cell' units. This method was particularly useful in studying speech processing [34], preserving and reconstructing multi-sentence paragraphs [35] and semantic representations [36].

(3) A gated recurrent unit (GRU) method was also chosen for the baseline [37]. GRU enabled each recurrent unit to adaptively capture dependencies of different time scales. Similar to the LSTM unit, the GRU had gating units that modulate the flow of information inside the unit, however, without having separate memory cells [38].

### 4. Results

### 4.1. Data pre-processing

Data pre-processing was carried out on both online job postings and the CGCO. After deleting unrelated job content, such as resume delivery mail, company address, non-Chinese texts and duplicated texts, a set of 107,328 data was retained. Regarding the CGCO, the ICTCLAS[39], was used to segment the words and to filter stop words and symbols in the texts.

### 4.2. Preliminary corpus construction

#### 4.2.1. The determination of p

The parameter p is used to balance the effects of the length of document in equation (1). In general, the choice of this parameter depends on the prior knowledge of the length of document. When containing more feature words, two texts that are not similar can be misjudged as similar texts. Considering words in two documents that lexical similarity is greater than P will reduce the influence effect of length of document.

To determine the parameter p in equation (1), 5000 job descriptions were randomly selected. For each one, the similarity values with the CGCO's 1481 job category description were computed. The category label with the largest value was regarded as the label of the job description. Then, five college students were asked to judge whether the label was correct. If the majority of people considered the label given by the computer correct, we concluded that the job description indeed belonged to that category in the CGCO. The test results for different p values are shown in Table 3. As shown in Table 3, WE-cos achieves its best results when p is 0.8.

*Table 3. Results of WE-cos on the determination of p.*

| Method | Parameter | Accuracy Rate |
|---|---|---|
| WE-cos | p=0.9 | 0.272 |
| WE-cos | p=0.8 | 0.302 |
| WE-cos | p=0.6 | 0.203 |
| WE-cos | p=0.4 | 0.174 |

*4.2.2. The preliminary corpus*

WE-cos with p=0.8 was used to process the input data to construct a preliminary corpus. The experimental process was the same as in the section 4.2.1. A set of 35546 job descriptions was correctly assigned with the CGCO's category labels. The result of data processing was described in Table 6.

Analysing the distribution of the job postings in the preliminary corpus, only 465 categories had data. This phenomenon is partly caused by the nature of online recruitment .Some occupations are not recruited through the internet, such as the leaders of a government, artists, and soldiers. Therefore, these categories were removed from our classification framework, resulting in 465 categories in JCTC.

## 4.3. Final corpus construction

When labelled data (the preliminary corpus) were ready, a classification method should be chosen to accelerate the construction of JCTC. Among the preliminary corpus, 70% of the texts were used as the training set and the rest as the test set. The hyper-parameter settings and the results of the RF classifier and SVM classifier are listed in Table 4 and Table 5, respectively.

The dictionary size refers to the number of feature words in the corpus. If all the words in the corpus are deemed as feature words, then the dictionary size is 15390. The dictionary size is 13381 and 8613 if the feature word occurrence is more than 3 times or 5 times, respectively.

*Table 4. Hyper-parameter tuning for RF classifier.*

| Number of Trees | 300 | 300 | 500 | 500 | 1000 |
|---|---|---|---|---|---|
| Size of dictionary | 8613 | 13381 | 13381 | 15390 | 13381 |
| Accuracy rate | 0.8369 | 0.8398 | 0.8381 | 0.8359 | 0.8356 |

*Table 5. Hyper-parameter tuning for SVM classifier.*

| Gamma | 0.006 | 0.006 | 0.006 | 0.06 | 0.6 | 0.7 |
|---|---|---|---|---|---|---|
| Size of dictionary | 8613 | 13381 | 15390 | 13381 | 13381 | 13381 |
| Accuracy rate | 0.7553 | 0.7554 | 0.7532 | 0.8808 | 0.8912 | 0.891 |

SVM outperforms the RF when SVM's gamma is 0.6 and dictionary size is 13381. Therefore, SVM with this hyper-parameter setting was used to process the unclassified data in Figure 1. The online job postings with labels unanimously agreed by SVM classifier and human judgement were added to the preliminary corpus. Afterwards, the corpus was used as training data to further process the unclassified data. This loop continued until the corpus no longer expanded. The final results of corpus expanding are shown in Table 6:

Table 6. Results of final corpus construction.

|         | Unclassified Data | Correct Classified Data | Remainder |
|---------|-------------------|-------------------------|-----------|
| WE-cos  | 107328            | 35127                   | 72201     |
| SVM-1st | 72201             | 64206                   | 7995      |
| SVM-2nd | 7995              | 3248                    | 4747      |

Table 6 demonstrates that 4747 job postings are unclassified, which are less than 5% of the total data. To verify these job postings are not able to be further added to JCTC, 100 samples from 4747 job posting were randomly selected. The scrutiny showed that most of them cannot be classified: 27 of them only contained advertisements about the company, 54 of them contained more than one job description. Considering the most of these data was useless, the 4747 remaining job postings were discarded.

Finally, a total of 102581 online job postings were correctly classified. They were distributed in 465 categories.

### 4.4. Comparison with other methods

Other methods mainly depend on human beings to directly classify texts to their corresponding category, such as [14][15]. This group of methods may work well when the number of category is small or the distinctions among categories are straightforward. However, it is not case for JCTC. JCTC has 465 categories. Besides, the distinctions among categories are sometimes vague as shown in Table 2.

To further validate the effectiveness of our method, 1000 samples were randomly selected from JCTC as a test set. These samples had "correct labels" assigned during the process of JCTC construction. Then, five college students were asked to assign one category label to each of these 1000 samples. For a certain sample, if the category label assigned by the majority of five students was same as the correct label, it was then defined as a correctly classified sample. Otherwise, it was named as an incorrectly classified sample. Uncertain sample meant no category label could get the majority vote. The experimental results are shown in Table 7.

Table 7. The results of human judgement

| # of total samples | # of correctly classified samples | # of incorrectly classified samples | # of uncertain samples | Accuracy rate |
|---|---|---|---|---|
| 1000 | 485 | 145 | 370 | 0.485 |

The number of uncertain samples accounts 37% of the population. This result suggests that subjective factors have a great influence on final performance. In our method, each label is given to a job posting if and only if the label assignment unanimously agreed by both the algorithm and humans. Besides, humans are only required to judge the correctness of the assigned labels rather than to classify job postings directly to categories. As a result, our method can reduce the subjectivity influence.

This experiment took the 5 students around 4 days. Each person worked at least 6 hours per day. However, the accuracy rate is only 48.5%. We inferred that the manual classifying would require more than 430 days if the data size was increased to 107382. By contrast, the method proposed here only used 50 days to address all the data with the 95.58% accuracy rate.

Hence, the method constructing JCTC reduces the influence of subjectivity and improves the efficiency of corpus construction as well.

## 4.5. Benchmark

Five state-of-art deep learning approaches were employed to validate the value of JCTC. The evaluation metric used was the accuracy rate. The experimental results are shown in Figure 3.

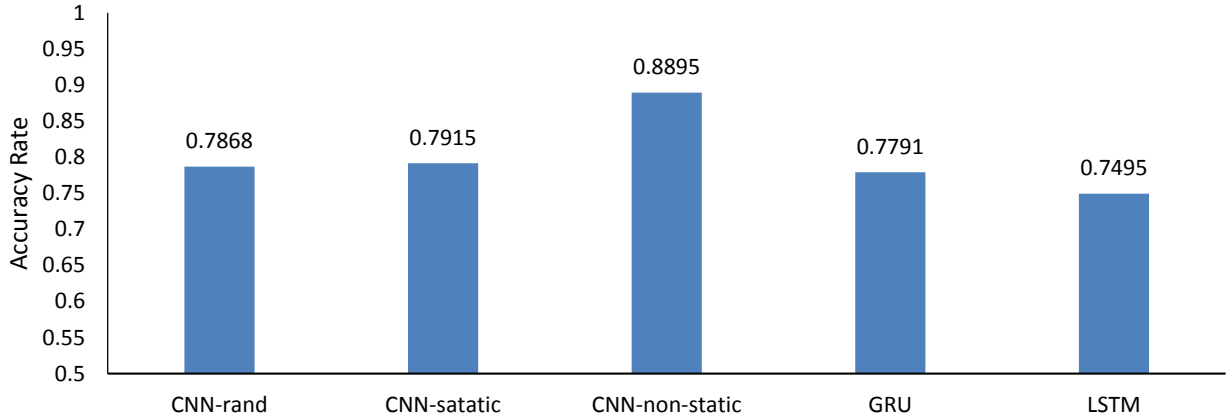

*Fig* 3. *The classification result on JCTC*

The experiments used 10-fold cross-validation. This allowed a proportion 90% of JCTC data to be used for training while making use of remaining data to access performance. For all CNN models, the activation functions used rectified linear units. The size of filter windows were 3, 4, 5 with 100 feature maps each.

During the training period, the min-batch size was 10 and the dropout rate equalled to 0.5. The Adadelta algorithm was used for weights parameter optimization and learning decay rate was set to be 0.95. The learning process was stopped after ~23K iterations. The hyper-parameters of the GRU approach were listed as follows: number of hidden units of 200, dropout rate of 0.5, min-batch size of 50, the maximum number of epoch of 15, and learning rate of 0.0004. For the case of LSTM, the hyper-parameters were number of hidden units of 200, dropout rate of 0.5, min-batch size of 50, the maximum number of epoch of 20, and learning rate of 0.0002. The word vectors fed into GRU and LSTM were pre-trained and fixed during the training period.

The results provide two insights about these approaches. Firstly, the CNN family outperforms the LSTM group. Even the CNN-rand having the poorest result among CNN family achieves a better performance than both GRU and LSTM. It indicates that CNN excels in classification while LSTM is good at sequence to sequence learning. Secondly, CNN-non-static has the best performance compared with other CNN approaches. This result suggests that the adaptive word embedding can provide further improvements.

## 5. Discussion

### 5.1. JCTC statistics

The most salient property of JCTC is that it is an imbalanced dataset. JCTC contains 102581 job postings, which spans in 465 categories. The number of job postings in each category varies widely, from 9632 job postings under "Salesman", to only 1 job posting under "Rail train driver". As shown in Table 8, most of the job postings belong to the top-level category called as "Professional and technical personnel". On the other hand, the amount of data in the "State organs, Party organizations, enterprises and institutions responsible person" level is the least, only accounting for 0.14% proportion of JCTC. This phenomenon is mainly caused by two reasons. One is different needs for various occupations in the labor market. Another is that occupations in certain categories are not recruited on the Internet because of their job natures.

*Table 8.* The statistics of JCTC

| Codes | Top-level category | Number of samples | Proportion |
| --- | --- | --- | --- |
| 1 | State organs, Party organizations, Enterprises and institutions responsible person | 141 | 0.14% |
| 2 | Professional and technical personnel | 66371 | 64.7% |
| 3 | Clerk and related personnel | 5240 | 5.11% |
| 4 | Business and service personnel | 27256 | 26.57% |
| 5 | Agricultural, Forestry, Animal husbandry, Fishery, Water conservancy Industry production personnel | 553 | 0.54% |
| 6 | Production, Transport equipment operators and related personnel | 3020 | 2.94% |

There are 50 forth-level categories whose number of job posting are over 500, and even 24 forth-level categories whose number of job posting are over 1000.

### 5.2. Comparison with other corpuses

Table 9 compares the JCTC with other text classification corpuses in terms of number of categories, number of samples, language, and source and hierarchy. Compared with other corpuses, JCTC has three advantages.

First, to the best of author's knowledge, JCTC is the first text classification corpus in the field of job market, which links online job information to the CGCO. JCTC would be beneficial for organizations which specialize in analysing the labor market or economy. By contrast, the sentiment classification corpuses [10][40][41] are generally built from movie reviews or product evaluations, and most of the other corpuses [9] [16] [42] come from news.

Second, JCTC is a text classification corpus containing the largest number of samples in Chinese. There are 24 categories having samples more than 1000. Although Fudan corpus is also a text classification corpus in Chinese[7], it has only 20 categories and contains less than 20 thousand samples.

Finally, JCTC is a hierarchical corpus, which means all categories are organized into a 4-level hierarchy analogy to the CGCO. Therefore, it can be used for hierarchical text classification which is a key technology for Large Scale text information organization with great values in the fields of the news and publishing, patent information management, etc. [43][44]. Although RCV1 is a hierarchical corpus, not all of the parent categories contain subcategories. The following lists a piece of the RCV1：

  CCAT (Corporate/Industrial)
    C14 (Share Listings)
    C15 (Performance)
      C151 (Accounts/Earnings)
        C1511 (Annual Results)
      C152 (Comment/Forecasts)
    C16 (Insolvency/Liquidity).

It shows that the category of C1511 is the only forth-level category under the top-level CCAT. Besides, most of the RCV1' data are classified in its secondary level. Therefore, The RCV1's structure is not sufficient for the hierarchical text classification.

*Table 9. Comparison with other text classification corpuses.*

| Corpus | Number of categories | Number of samples | Language | Source | Hierarchy |
|---|---|---|---|---|---|
| 20News[9] | 20 | 19997 | English | News | N |
| RCV-1[16] | 103 | 804414 | English | News | Y |
| Reuters-21578[42] | 118 | 12902 | English | News | N |
| Fudan [7] | 20 | 19636 | Chinese | News | N |
| SST-1[40] | 5 | 11855 | English | Movie reviews | N |
| SST-2[40] | 2 | 9613 | English | Movie reviews | N |
| MR[10] | 2 | 10662 | English | Movie reviews | N |
| CR[41] | 2 | 3775 | English | Product evaluations | N |
| JCTC | 465 | 102581 | Chinese | Recruitment information | Y |

## 6. Conclusions

The absence of the appropriate recruitment information corpus makes the massive amount of online job information unusable for the labor market analysis and prediction. On one hand, prior text classification corpuses mainly focus on news, movie reviews or product evaluations, such as [9], [10], [11], etc. On the other hand, some recruitment information databases are lack of a uniform job title definition [2].

This papers introduces the HR-CTCJCTC involving 102581 online job postings originally crawled from three mainstream recruitment websites in China. JCTC contains 465 categories. In JCTC construction framework, a formal specification issued by the Chinese central government is chosen as the classification standard. Then, a preliminary corpus is built through the unanimous agreement between the unsupervised learning algorithm and human judgement. Afterwards, the corpus continues to expand with the help of the preliminary corpus acting as the training dataset. The job postings would further be added to the corpus if they are simultaneously approved by the supervised learning approach and human judgement. The JCTC is the final corpus when no more job postings could be added in.

The method used in constructing of JCTC has several merits. Firstly, it "downgrades" the human role in the construction process. It does employ human judgement just like other existing methods. Nevertheless, people are only required to determine whether the job postings should be classified in certain categories or not, while other methods would ask people to directly classify data into categories. The method proposed here can not only ameliorate the high demands on people's skill and knowledge, but reduce the subjective influence as well. Secondly, the method is not limited to Chinese. It can be applied to other languages considering that most nations have specifications similar to the CGCO.

JCTC, as an organized dataset of job postings, fills the gap in text classification corpus realm. Existing corpuses mainly focus on movie reviews, news and product evaluations. The massive amount of online recruitment information seems to be neglected in this big data era. An appropriate corpus is a fundamental building block to push forward research in this field, although corpus construction itself seems not to have lots of academic innovation.

JCTC has the largest number of samples among text classification corpuses in Chinese. The ICTCAS's Chinese news corpus [6] only consists of 16280 samples. Fudan corpus [7] only contains 9833 samples in test corpus and 9804 samples in training corpus. TanCorpV1.0. [8] contains only 14150 samples.

JCTC has a profound implications in text classification community, professionals who pay close attention to macroeconomics and even the statistical authorities. This paper provides benchmark on JCTC by five state-of-the-art deep learning approaches. Future algorithms will be developed with the help of the benchmark. Besides, JCTC borrows the classification system from the CGCO. Hence it is inherently hierarchical corpus that could be applied in hierarchical text classification research. On the other hand, professionals will be able to use the plentiful online job data to analyze the labor market, or even the national economy with the help of the JCTC. It took the statistical authorities 5 years and thousands of experts from nearly 70 industries to compile the CGCO used in this paper. The authorities are now armed with a novel tool to monitor the newly emerged occupations on the Internet. Afterwards, they can update the CGCO in a more efficient and effective manner.

We will continue to expand the corpus and make it publicly available for the research community. It is hopeful that JCTC will encourage NLP practitioners to pay more attention to the dataset construction mechanism in the future. Although corpus construction is a labor intensive task, it is an elementary impetus to the NLP research just like fuels to the rockets. Meanwhile, it would be more delightful if JCTC assists any macroeconomists to perform comprehensive, accurate and timely analysis on the job market.

*Acknowledgments*

This research has been partially co-financed by strategic pilot technology Chinese Academy of Sciences (Grant No. XDA06010301) , Project of International Corporation and Exchange NSFC (Grant No. 61461136001), and Zhang Jiang high tech Park Management Committee(Grant No.2016-14).